\documentclass[a4paper,11pt]{article}

\usepackage{jheppub} 

\usepackage[latin1]{inputenc}
\usepackage{amsmath}
\usepackage{slashed}
\usepackage{graphicx}
\usepackage{verbatim}
\usepackage{multirow}
\usepackage[labelfont=bf,textfont={sl,bf}]{subfig}

\newcommand{\GeV}{{\,\mathrm{GeV}}}
\newcommand{\TeV}{\,\mathrm{TeV}}

\newcommand{\fb}{{\,\mathrm{fb}}}

\newcommand{\order}[1]{\mathcal{O}\!\left(#1\right)}

\newcommand{\bea}{\begin{eqnarray}}
\newcommand{\eea}{\end{eqnarray}}

\newcommand{\crn}{\nonumber \\}

\newcommand{\eq}[1]{Eq.~(\ref{#1})}
\newcommand{\bib}[1]{Ref.~\cite{#1}}
\newcommand{\fig}[1]{Fig.~\ref{#1}}
\newcommand{\tab}[1]{Table~\ref{#1}}
\newcommand{\sect}[1]{Section~\ref{#1}}

\newcommand{\appen}[1]{Appendix~\ref{#1}}

\title{Next-to-leading order QCD corrections to $ZZ$ production
  in association with two jets}

\preprint{FTUV-14-0329\;\;IFIC-14-31\;\;KA-TP-12-2014\;\;
  \\ \hspace*{8.75cm} LPN14-070\;\; SFB/CPP-14-25}

\author[a]{Francisco Campanario,}
\author[b]{Matthias Kerner,}
\author[b,c]{Le Duc Ninh}
\author[b]{and Dieter Zeppenfeld}

\affiliation[a]{Theory Division, IFIC, University of Valencia-CSIC, E-46980
  Paterna, Valencia, Spain}
\affiliation[b]{Institut f\"ur Theoretische Physik, Karlsruher Institut f\"ur  Technologie, \\
D-76128 Karlsruhe, Germany}
\affiliation[c]{Institute of Physics, Vietnam Academy of Science and Technology, \\
10 Dao Tan, Ba Dinh, Hanoi, Vietnam}

\emailAdd{francisco.campanario@ific.uv.es}
\emailAdd{matthias.kerner@kit.edu}
\emailAdd{duc.le@kit.edu}
\emailAdd{dieter.zeppenfeld@kit.edu}

\abstract{We present the first calculation of 
next-to-leading order QCD corrections to QCD-induced $ZZ$ production in association with two jets at 
hadron colliders.  
Both $Z$ bosons decay leptonically with all off-shell effects, virtual photon contributions and spin-correlation effects 
fully taken into account. 
This process is an important background to weak boson scattering, to the
measurement of quartic gauge couplings and to searches for signals of new physics beyond the Standard Model. 
As expected, the next-to-leading order corrections reduce significantly the scale
uncertainty and show a non-trivial phase space dependence in kinematic distributions. 
Our code will be publicly available as part of the parton
level Monte Carlo program {\texttt{VBFNLO}}. 
}


\begin{document}
\maketitle
\flushbottom

\section{Introduction}
\label{intro}

With the first measurement~\cite{ATLAS001} of same-sign $W^\pm W^\pm jj$ vector boson production 
at the LHC ATLAS experiment at $\sqrt{s} = 8\TeV$, the program to test $VV \to VV$ scattering and 
the EW quartic gauge couplings has been started. The data are in agreement with the Standard Model (SM)
prediction~\cite{Jager:2009xx,Denner:2012dz,Melia:2010bm,Campanario:2013gea}
and provide the first evidence for electroweak (EW) gauge boson scattering, namely $W^\pm
W^\pm \to W^\pm W^\pm$. 
In this context, the class of processes with two EW gauge bosons and two jets in the final state plays a very 
important role. Furthermore, these processes are important backgrounds for
searching signals of new physics beyond the SM. 

From the theoretical side, progress has been made to provide predictions at next-to-leading order (NLO) QCD accuracy. 
The strategy used so far is first to implement the calculation of the hard processes 
\bea
pp \to VV jj + X \label{eq:ppVVjj process}, 
\eea
where both gauge bosons decay leptonically, in a parton
level Monte Carlo program, where parton distribution functions and a jet algorithm 
to cluster final state partons into jets are applied, and then interface it to 
other programs which can do parton shower and hadronization. 
From a physical point of view and also due to the complexity of the calculation, it has 
been traditional to classify the process, \eq{eq:ppVVjj process}, into EW and QCD induced 
contributions based on the difference in the overall coupling constant at leading order (LO) and 
calculate them separately. The interference effects between these two contributions are 
expected to be negligible for most measurements at the LHC. However, if needed,  
one can calculate these effects at LO using an automated 
program e.g. {\texttt{Sherpa}}~\cite{Gleisberg:2008ta}. For a recent discussion on 
this issue, we refer to \bib{Campanario:2013gea}, where the interferences of
the same-sign $W^\pm W^\pm jj$ vector boson production process were studied, which are
expected to be maximal because the 
gluon-initiated subprocesses are absent at LO and both 
the EW and QCD amplitudes involve only left-chiral quarks and leptons. 

The EW-induced channels of order $\order{\alpha^4}$ for 
on-shell production at LO are further 
classified into ``vector boson fusion'' (VBF) mechanisms, 
which are sensitive to the EW quartic gauge couplings and the dynamics of $VV \to VV$ scattering and other 
contributions including $VVV$ production with 
one $V$ decaying into two jets. The important message is that the VBF mechanisms 
can be strongly enhanced if VBF cuts are applied. References for the NLO QCD calculations of the 
EW-induced channels and the definition of the VBF cuts can be found in \bib{Campanario:2013gea}. 

In this paper, we consider the $\order{\alpha^2\alpha_s^2}$ QCD-induced mechanism for the process with $ZZjj$ in the final state and 
will present the first theoretical prediction at NLO QCD accuracy. 
The NLO QCD computation of the corresponding EW-induced VBF mechanism has been done in \bib{Jager:2006cp}. 
The NLO QCD corrections to the QCD-induced channels 
are much more difficult because QCD 
radiation occurs already at LO, leading to complicated topologies (up to hexagons with rank-5 tensor integrals) 
with non-trivial color structures at NLO. The calculations for $W^+W^-jj$ production have been presented in 
Refs.~\cite{Melia:2011dw, Greiner:2012im}, for the same-sign $W^+W^+jj$ in 
Refs.~\cite{Melia:2010bm,Campanario:2013gea} and for $W^\pm Zjj$ in \bib{Campanario:2013qba}. 
Similar calculations with the massless photon in the final state have also been calculated  
for $\gamma \gamma jj$~\cite{Gehrmann:2013bga,Badger:2013ava,Bern:2014vza}
and $W \gamma jj$~\cite{Campanario:2014dpa} production. 
Results for $\gamma \gamma jjj$ 
production at NLO QCD have been very recently presented also in
Ref.~\cite{Badger:2013ava}. Results at the total cross section level for
on-shell $VVjj$ production have been very briefly reported recently in Ref.~\cite{Alwall:2014hca}.

Our $ZZjj$ calculation with leptonic decays 
has been implemented within the {\texttt{VBFNLO}}
framework~\cite{Arnold:2008rz,Baglio:2014uba}, a parton level Monte Carlo program
which allows the definition of general acceptance cuts and
distributions. As customary in {\texttt{VBFNLO}}, 
all off-shell effects, virtual photon contributions and spin-correlation effects 
are fully taken into account. In this paper, we focus on the four charged-lepton final states. 
The $l^+l^-\bar{\nu}\nu$ channels are simpler and can be easily adapted from the 
four charged-lepton code (e.g. switching off a virtual photon 
contribution, changing the lepton-Z couplings). 
This possibility will be available in the next release of {\texttt{VBFNLO}}. 

The outline of this paper is the following. Details of our calculation 
are provided in \sect{sec:calculation}. 
In \sect{sec:results}  
numerical results for inclusive cross sections and various distributions are given. 
Conclusions are presented in \sect{sec:con} and in \appen{sec:appendixA} 
results at the amplitude squared level for a random phase-space point are
provided in order to facilitate comparison with independent calculations.
\section{Calculational details}
\label{sec:calculation}
In this paper, we calculate the QCD-induced processes at NLO QCD for
the process
\begin{equation}
\label{eq:process}
pp \to l_1^+ l_1^- l_2^+ l_2^- j j + X,
\end{equation}
at order ${\cal O}(\alpha_s^3 \alpha^4)$. We assume that all the leptons are massless and $l_1 \neq l_2$. 
\begin{figure}[h]
  \centering
\includegraphics[width=0.45\textwidth]{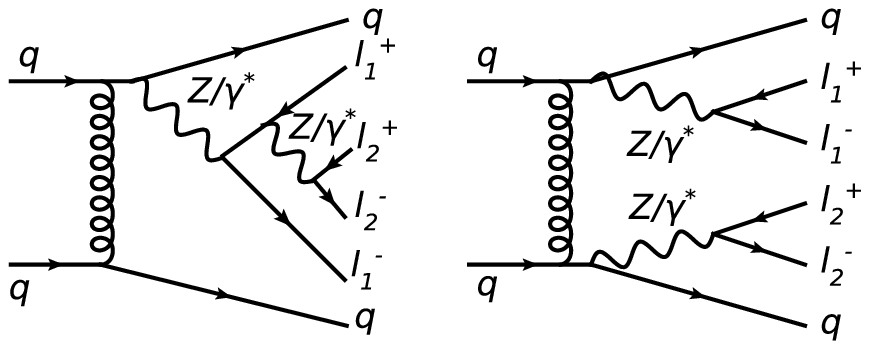}
\includegraphics[width=0.45\textwidth]{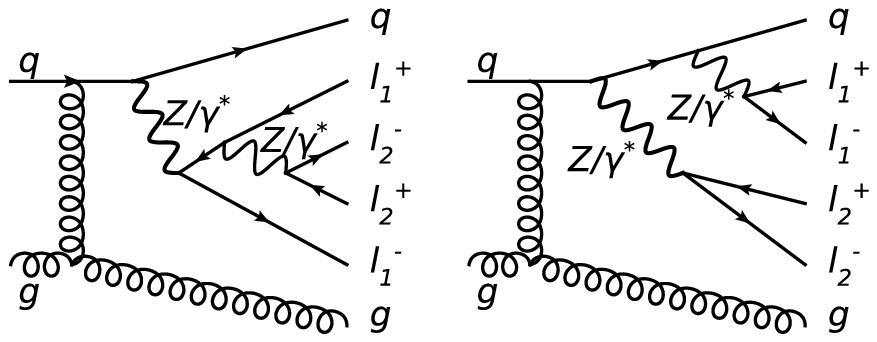}
\caption{Representative tree-level Feynman diagrams.}
\label{fig:feynTree}
\end{figure}
This process is then divided into the contributions 
\begin{align}
pp &\to \hat{V} j j + X, \label{eq:processVhat}\\
pp &\to V_1 V_2 j j + X \label{eq:processVV},
\end{align}
where $\hat{V} = Z,\gamma^*$ 
with subsequent decay $\hat{V} \to l_1^+ l_1^- l_2^+ l_2^-$ 
and $V_i = Z,\gamma^*$ with $V_i \to l_i^+ l_i^-$ ($i = 1,2$). 
Some representative Feynman diagrams are shown in \fig{fig:feynTree}. 
At the LHC, the dominant contribution comes from the 
process in \eq{eq:processVV} with $V_1 = V_2 = Z$ because 
the gauge bosons can be both simultaneously on-shell. 
For simplicity, we describe the resonating $Z$ propagators with a fixed width and
keep the weak-mixing angle real. Moreover, 
since those leptonic decays of the neutral gauge bosons are consistently 
included in our calculation by the replacement of the polarization vectors 
with the corresponding effective currents, we will sometimes refer to the process 
in \eq{eq:process} as $ZZjj$ production. 

We use the Feynman
diagrammatic approach and classify at LO the above contributions
into $4$-quark and $2$-quark-$2$-gluon amplitudes
\begin{align}
uu &\to uu\; V_1 V_2,\crn
uc &\to uc\; V_1 V_2,\crn
ud &\to ud\; V_1 V_2,\crn
dd &\to dd\; V_1 V_2,\crn
ds &\to ds\; V_1 V_2,\crn
gg &\to \bar{u}u\; V_1 V_2,\crn 
gg &\to \bar{d}d\; V_1 V_2,
\label{eq:subproc}
\end{align}
where the sub-dominant processes in \eq{eq:processVhat} and the leptonic decays 
are implicitly included. 

From these seven generic subprocesses we 
can obtain all the amplitudes of other subprocesses via crossing or/and exchanging the partons. 
We work in the 5-flavor scheme, i.e. external bottom-quark contributions with $m_b = 0$ are included. 
Subprocesses with external top quarks are excluded, 
but virtual top-loop contributions are included in our calculation as specified below. 

At NLO QCD, there are the virtual and the real-emission 
corrections. We use dimensional regularization~\cite{'tHooft:1972fi}
to regularize the ultraviolet (UV) and infrared (IR) divergences and
use an anticommuting prescription of
$\gamma_5$~\cite{Chanowitz:1979zu}. The UV divergences of the virtual
amplitude are removed by the renormalization of $\alpha_s$. Both the
virtual and the real corrections are infrared divergent. These
divergences are canceled using the Catani-Seymour prescription~\cite{Catani:1996vz} such
that the virtual and real corrections become separately numerically
integrable. 
The real emission contribution
includes, allowing for external bottom quarks, $275$ subprocesses with seven
particles in the final state. 
\begin{figure}[h]
  \centering
\includegraphics[scale=0.85]{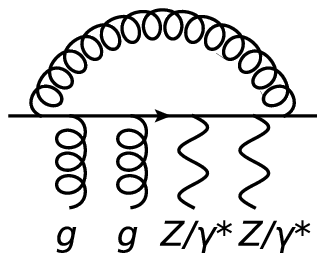}   \hfill 
\includegraphics[scale=0.85]{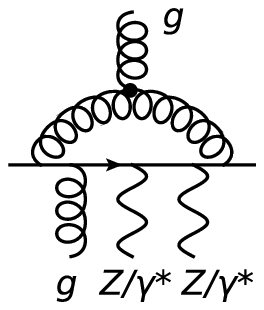} \hfill
\includegraphics[scale=0.85]{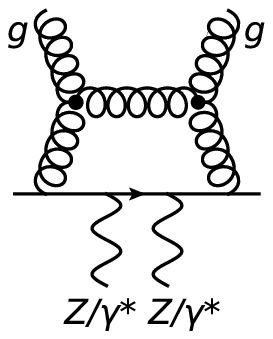}    \hfill
\includegraphics[scale=0.85]{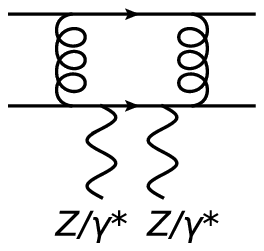}    \hfill
\includegraphics[scale=0.85]{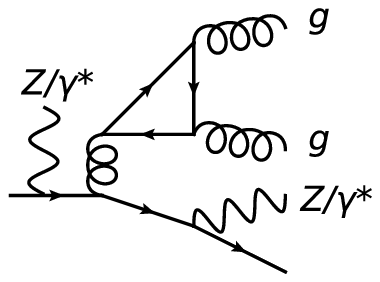}     
\caption{Selected Feynman diagrams contributing to the virtual amplitudes.}
\label{fig:feynVirt}
\end{figure}

The virtual amplitudes are more
challenging involving up to six-point rank-five one-loop
tensor integrals appearing in the $2$-quark-$2$-gluon virtual
amplitudes. There are $42$ six-point diagrams for 
the last subprocess in \eq{eq:subproc}. Since the kinematics does not change if a $Z$ boson is 
replaced by a virtual photon, the same hexagon integrals occur 
in the $ZZ$, $Z\gamma^*$, $\gamma^* Z$ and $\gamma^* \gamma^*$ contributions 
and hence are reused for optimization. 
The $4$-quark group is simpler with $24$ generic hexagons for
the first subprocess in \eq{eq:subproc}. For the other $4$-quark subprocesses with different 
flavors there are $12$ hexagons. 

In \fig{fig:feynVirt}, we highlight some contributions to the virtual
amplitude. 
Diagrams including a closed-quark loop with gluons attached to it, e.g. the last diagram of \fig{fig:feynVirt}, 
are taken into account. 
However, we do not include closed-quark loops
where the $Z/\gamma^*$ vector bosons or/and the Higgs boson are directly attached to them. This set of diagrams forms a gauge
invariant subset and contributes at the few per mille
level to the NLO results~\cite{Greiner:2012im}, and hence are negligible for all phenomenological
purposes. On the other hand, the discarded diagrams, which include the loop-induced 
$gg \to H(\to ZZ)jj$ channels, can be regarded as a new mechanism, which also receives contributions from 
$gg \to ZZ gg$. To properly take into account these loop-induced channels one has to 
calculate the square of those amplitudes, which formally are of higher-order but can be somewhat enhanced by the 
Higgs resonance and the large gluon luminosity at the LHC. 
Part of those amplitudes can be obtained from the $gg \to ZZg$ calculation presented in \bib{Campanario:2012bh}. 
This effect is expected to be at the few percent level, which is of 
similar size as the interferences between the EW and QCD induced mechanisms discussed in the introduction. 
We note that NLO EW corrections are also at the same level.


The evaluation of scalar integrals is done following 
Refs.~\cite{'tHooft:1978xw, Bern:1993kr,Dittmaier:2003bc,Nhung:2009pm,Denner:2010tr}. 
The tensor coefficients of the loop 
integrals are computed using the Passarino-Veltman reduction
formalism~\cite{Passarino:1978jh} up to the box level. 
For pentagons and hexagons, 
we use the reduction formalism of Ref.~\cite{Denner:2005nn} (see also
Refs.~\cite{Campanario:2011cs,Binoth:2005ff}).

Our calculation has been carefully checked as follows. 
The present calculation shares a large common part with our 
previous $W^\pm Zjj$ calculation~\cite{Campanario:2013gea}, which has been 
validated at the amplitude level performing two independent calculations. 
For the real-emission part, the structure of QCD radiation and therefore 
the implementation of the Catani-Seymour dipole subtraction method is the same. 
The only difference is the computation of the tree-level amplitudes. 
Using two independent codes, we have crosschecked the real-emission amplitudes and the corresponding 
subtraction terms at a random phase-space point and obtained 10 digit agreement 
with double precision. Similarly, the integrated part of the 
dipole subtraction term defined in~\bib{Catani:1996vz} has been validated at the integration level. 
Moreover, the real-emission contribution including the subtraction terms has been crosschecked
against Sherpa~\cite{Gleisberg:2008ta,Gleisberg:2008fv} and agreement at the per mill level was found. 
For the virtual part, 
we have again checked the whole virtual amplitudes with two independent 
calculations and obtained full agreement, typically $6$ to $12$ digits 
with double precision, at the amplitude level. 
The first implementation uses
{\texttt{FeynArts-3.4}}~\cite{Hahn:2000kx} and
{\texttt{FormCalc-6.2}} \cite{Hahn:1998yk} to obtain the virtual
amplitudes. The in-house library {\texttt{LoopInts}} is used to
evaluate the scalar and tensor one-loop integrals. 


For the second implementation, which will be
publicly available via the {\texttt{VBFNLO}} program and is the one 
used to obtain the numerical results presented in the next section, 
we use the in-house library presented in Ref.~\cite{Campanario:2011cs} to
compute the amplitudes and evaluate the tensor integrals. 


Furthermore, we closely follow the strategy described in
 Refs~\cite{Campanario:2011cs,Campanario:2012bh,Campanario:2013gea,Campanario:2014dpa} 
to optimize the code and to deal with numerical instabilities occurring in the 
numerical evaluation of the virtual part. 
With this method, we obtain the NLO inclusive cross section with
statistical error of $1\%$ in 3.5 hours on an Intel $i5$-$3470$
computer with one core and using the compiler Intel-ifort version
$12.1.0$. The distributions shown below are based on multiprocessor runs
with a total statistical error of 0.03\%.

\section{Numerical results}
\label{sec:results}
As input parameters, we use  $M_W=80.385 \GeV$, $M_Z=91.1876 \GeV$, 
$m_t=173.1\GeV$ and
$G_F=1.16637\times 10^{-5}\GeV^{-2}$. The tree-level relations 
are then used to calculate the weak mixing angle and the 
electromagnetic coupling constant. We use the MSTW2008 parton
distribution functions~\cite{Martin:2009iq} with
$\alpha_s^\text{LO}(M_Z)=0.13939$ and
$\alpha_s^\text{NLO}(M_Z)=0.12018$. 
The $Z$ total width is calculated as $\Gamma_Z = 2.508905\GeV$. 
All fermions but the top quark are approximated as massless. 
We work in the five-flavor scheme and use the 
$\overline{MS}$ renormalization of the strong coupling constant with the top quark decoupled from the running of $\alpha_s$. 
However, the top-loop contribution is
explicitly included in the virtual amplitudes. We choose inclusive cuts defined as 
\begin{align}
  p_{T(j,l)} &> 20 \GeV&  |y_j| &< 4.5& \crn
  |y_l| &< 2.5&  R_{l(l,j)} &> 0.4,&
\end{align}
where the anti-$k_t$
algorithm~\cite{Cacciari:2008gp} with a cone radius of $R=0.4$ is used
to cluster partons into jets. 
For the cut on $R_{lj}$, all reconstructed jets are taken into account. 
We use a dynamical factorization and renormalization scale with the central value
\bea
\mu_{F} = \mu_{R} = \mu_{0}=\frac{1}{2}
\left[E_{T}(jj) + E_{T}(4l)\right],
\label{eq:define_mu0}
\eea
where $E_{T} = (p_T^2 + p^2)^{1/2}$ is calculated for
the systems of the two tagging jets and of the four leptons. 
The two tagging jets are defined as the highest transverse-momentum jets. 
This scale choice is well motivated because the $E_{T}(jj)$ term interpolates between 
the transverse momenta and the invariant mass of the tagging-jet system. 
It is therefore similar to the 
default scale defined in \bib{Campanario:2014dpa}. We have checked that 
the two scale choices indeed produce nearly identical results for various 
kinematic distributions at both LO and NLO levels. 
In the following, results for the integrated cross section and
for various differential distributions with the above setting will be 
presented. We sum over all possible combinations of charged leptons of 
the first two generations, i.e. final states $e^+e^-\mu^+\mu^-$, $e^+e^-e^+e^-$ 
and $\mu^+\mu^-\mu^+\mu^-$ are all included. Since Pauli-interference effects for 
the identical lepton channels are neglected, this sum amounts to a factor of two 
compared to the single $e^+e^-\mu^+\mu^-$ result. 

\begin{figure}[h!]
  \centering
  \includegraphics[width=0.6\textwidth]{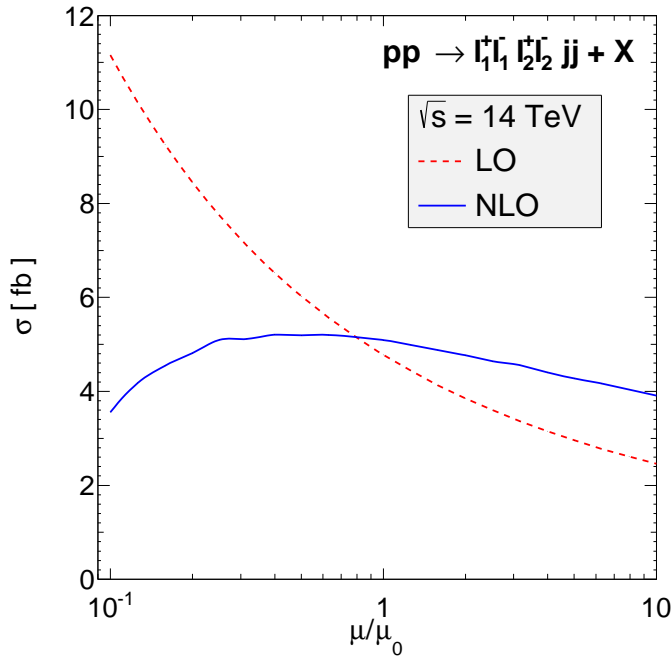}
\caption{Scale dependence of the LO and NLO cross sections at the LHC. 
The reference scale $\mu_0$ is defined in \eq{eq:define_mu0}. 
All possible combinations of charged leptons of the first two generations 
are included.}
\label{fig:scale}
\end{figure}
As customary in the framework of perturbative QCD, our NLO results depend on 
the scales $\mu_F$ and $\mu_R$. We set them equal for simplicity. 
The scale dependence of the total cross section at LO and NLO is shown in 
\fig{fig:scale}. At the default scale $\mu_0$, we obtain 
$\sigma_\text{LO} = 4.7783(3)^{+1.25}_{-0.93} \fb$ and 
$\sigma_\text{NLO} = 5.075(2)^{+0.13}_{-0.30} \fb$
where the numbers in the parentheses are the statistical errors of the numerical integrations 
and the other uncertainties are 
due to $\mu_0/2 \le \mu_F=\mu_R\le 2\mu_0$ variations. As expected, 
the scale dependence around the central value $\mu_0$ is significantly reduced 
when the NLO contribution is included. 

\begin{figure}[ht!]
  \centering
  \includegraphics[width=0.45\textwidth]{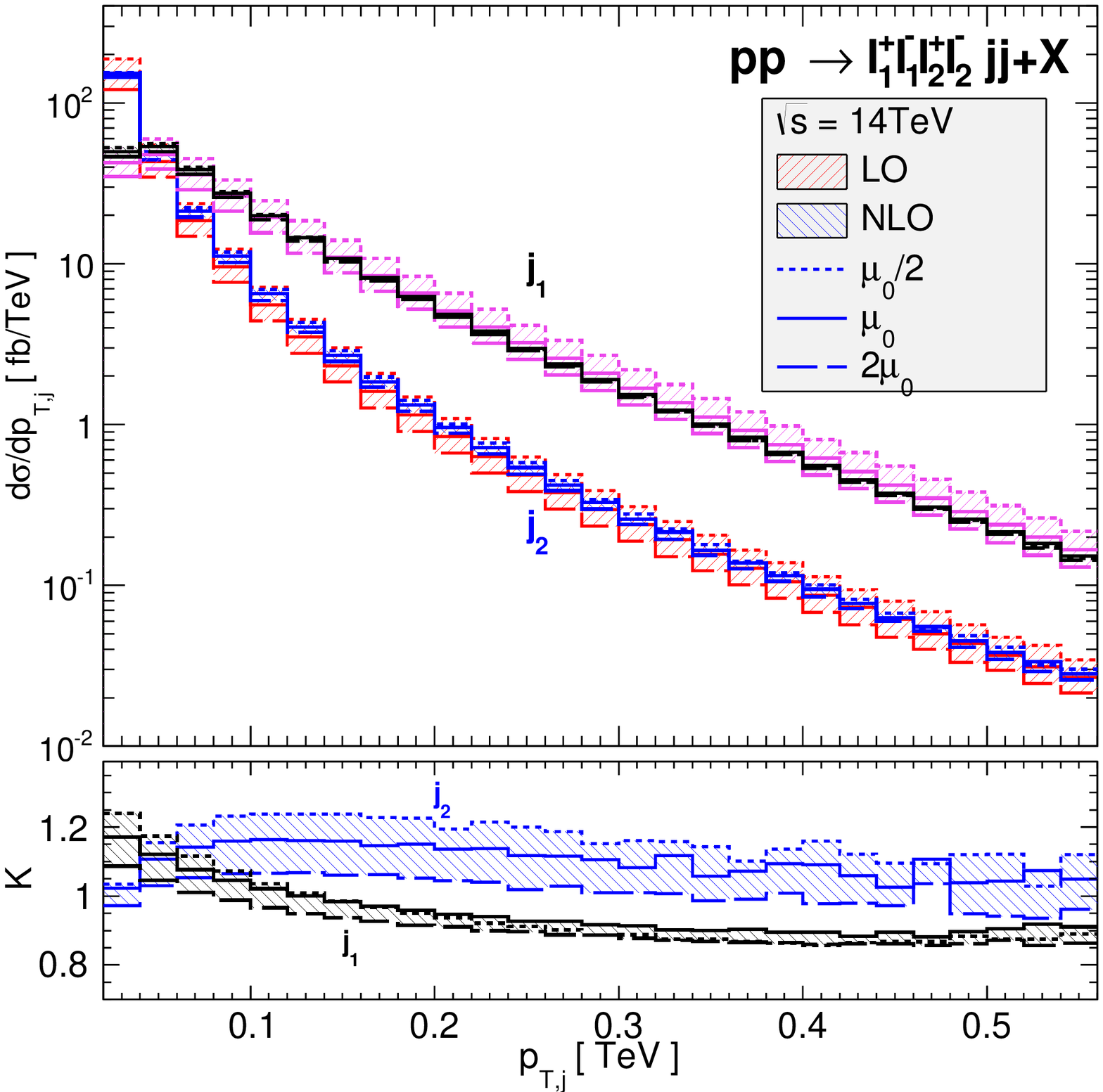}
  \includegraphics[width=0.45\textwidth]{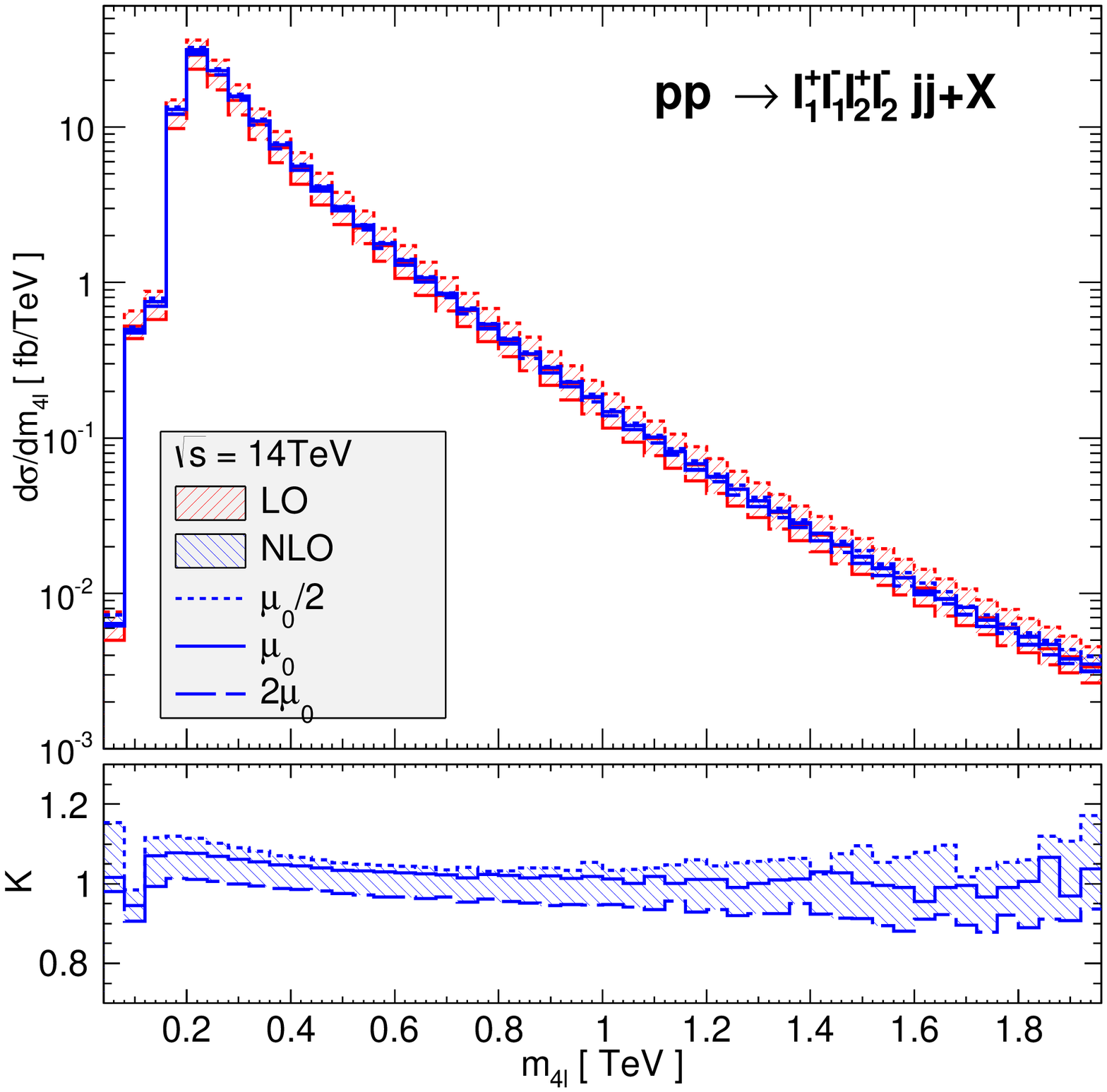}
  \includegraphics[width=0.45\textwidth]{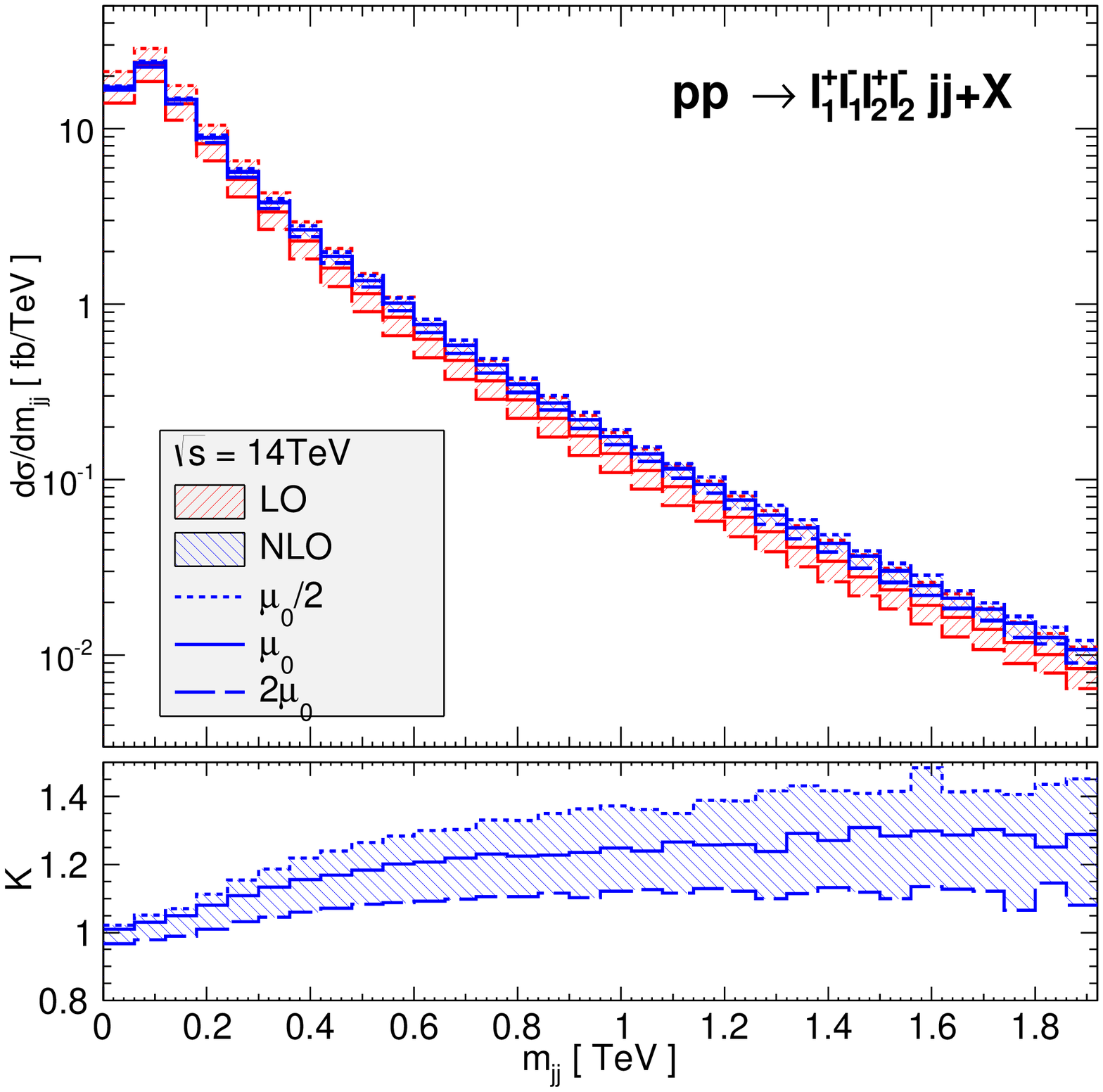}
  \includegraphics[width=0.45\textwidth]{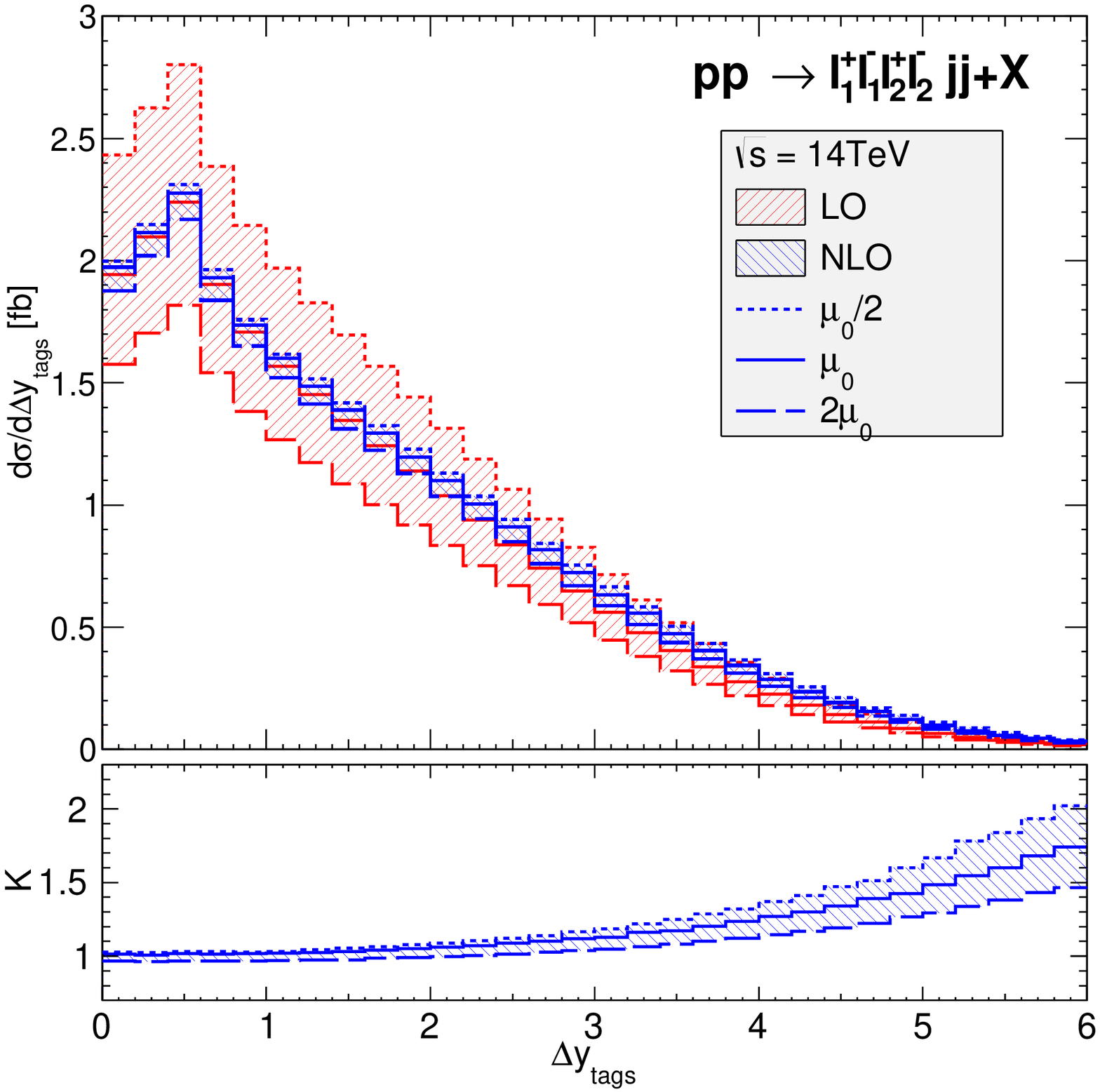}
  \caption{Differential cross sections 
  with inclusive cuts for the transverse momenta (top left),  
the invariant mass of the four-lepton system (top right) and 
of the two tagging jets (bottom left). 
The distributions of the rapidity separation between the two jets are in the bottom right panel. 
The bands describe $\mu_0/2 \le \mu_F=\mu_R\le 2\mu_0$ variations. 
The $K$-factor bands are due to the scale variations of the NLO results, 
with respect to $\sigma_\text{LO}(\mu_0)$. 
The solid lines are for the central scale while the dotted and dashed lines correspond to $\mu_0/2$ and $2\mu_0$, respectively.
All possible combinations of charged leptons of the first two generations 
are included.
}
\label{dist_NLO_jets_inc}
\end{figure}
We now study the phase space dependence of the NLO QCD corrections. 
In \fig{dist_NLO_jets_inc}, 
we display the distributions of the 
transverse momenta (top left) and 
the invariant mass (bottom left) of the two hardest jets and 
the invariant mass of the four-lepton system (top right). 
The distributions of the rapidity separation between the two jets are in the bottom right panel. 
The $K$ factors, defined as the ratio of the NLO to the LO results, are 
shown in the small panels. 
To give a measure of scale uncertainty, we vary the scales in the range 
$\mu_0/2 \le \mu_F=\mu_R\le 2\mu_0$ and plot the LO and NLO bands in the large 
panels. 
The $K$-factor bands are due to the scale variations of the NLO results, 
with respect to $\sigma_\text{LO}(\mu_0)$. 
As expected, the scale uncertainties for all the distributions are significantly 
reduced at NLO. We observe non-trivial behaviors of 
the $K$ factors, varying from $0.8$ to $1.7$ for the default scale. 
The rapidity-separation distribution shows that NLO QCD corrections are important 
at large separation ($\Delta y_\text{tags} > 4$), which is the phase space region 
selected by VBF cuts~\cite{Jager:2006cp} to enhance $VV\to VV$ scatterings. 

\section{Conclusions}
\label{sec:con}
In this paper, we have presented first results at NLO QCD for the four charged-lepton 
$l_1^+l_1^-l_2^+l_2^-$ production in association with two jets at the LHC via QCD-induced 
mechanisms at order ${\cal O}(\alpha_s^3 \alpha^4)$. 
The final-state leptons are created via a virtual photon or a $Z$ boson. 
The dominant contribution comes from the 
phase space regions where two intermediate $Z$ bosons are simultaneously resonant, 
therefore this process is usually referred to as $ZZjj$ production. 
All off-shell effects, virtual photon contributions and spin-correlation effects 
are fully taken into account. 
We have shown that the NLO QCD corrections are important and hence should be taken 
into account for precise measurements at the LHC. 
With this result the calculation of NLO QCD corrections 
to the production of two massive gauge bosons together with two jets is now practically complete. 

Our code will be publicly available as part of the {\texttt{VBFNLO}} 
program~\cite{Arnold:2008rz,Baglio:2014uba}, 
thereby further studies of the QCD corrections with different kinematic cuts can be easily done.

\appendix
\section{Results at one phase-space point}
\label{sec:appendixA}
In this appendix, results at a random phase-space point are 
provided for comparison with future independent calculations. We focus on the virtual
amplitudes, which are most complicated, of the seven benchmark subprocesses in \eq{eq:subproc}.
The amplitudes of all other subprocesses can be obtained via crossing or/and 
exchanging the partons. 
The phase-space point for the process $j_1 j_2 \to j_3 j_4 e^+ e^- \mu^+ \mu^-$ 
is given in \tab{table_PSP_2to6}, which is the same as the one in \bib{Campanario:2013gea}.
\begin{table}[th]
 \begin{footnotesize}
 \begin{center}
 \caption{\label{table_PSP_2to6}{Momenta (in GeV) at a random phase-space point for $j_1 j_2 \to j_3 j_4 e^+ e^- \mu^+ \mu^-$ subprocesses.}}
\begin{tabular}{l | r@{.}l r@{.}l r@{.}l r@{.}l}
& \multicolumn{2}{c}{ $E$}
& \multicolumn{2}{c}{ $p_x$}
& \multicolumn{2}{c}{ $p_y$}
& \multicolumn{2}{c}{ $p_z$}
\\
\hline
$j_1$  & 18&3459102072588 & 0&0  & 0&0  &  18&3459102072588  \\
$j_2$  & 4853&43796816526 & 0&0  & 0&0  &  -4853&43796816526  \\
$j_3$  & 235&795970274883 & -57&9468743482139 & -7&096445419113396$\times 10^{-15}$ & -228&564869022223 \\
$j_4$  & 141&477229270568 & -45&5048903376581 & -65&9221967646567 & -116&616359620580 \\
$e^+$  & 276&004829895761 & 31&4878768361538 & -8&65306166938040 & -274&066240646098 \\
$e^-$  & 1909&28515244344 & 29&6334571080402 & 40&1409467910328 & -1908&63311192893 \\
$\mu^+$  & 2241&46026948104 & 28&1723094714198 & 30&2470561132914 & -2241&07910976778 \\
$\mu^-$  & 67&7604270068059 & 14&1581212702582 & 4&18725552971283 & -66&1323669723852 \\
\hline
\end{tabular}\end{center}
 \end{footnotesize}
\end{table}
\begin{table}[th]
 \begin{footnotesize}
 \begin{center}
\caption{\label{table_PSP_QCD_jjjj}{QCD interference amplitudes $2\text{Re}(\mathcal{A}_\text{NLO}\mathcal{A}^{*}_\text{LO})$
for $j_1 j_2 \to j_3 j_4 e^+ e^- \mu^+ \mu^-$ subprocess.}}
\begin{tabular}{l | r@{.}l r@{.}l r@{.}l}
\hline
& \multicolumn{2}{c}{ $1/\epsilon^2$}
& \multicolumn{2}{c}{ $1/\epsilon$}
& \multicolumn{2}{c}{ finite}
\\
\hline
& \multicolumn{2}{c}{ }
& \multicolumn{2}{c}{ $uu \to uu$}
& \multicolumn{2}{c}{ }
\\
\hline
I operator & 2&8915745669$\times 10^{-6}$ & -1&4951973738$\times 10^{-6}$ & 6&947191076$\times 10^{-7}$ \\
loop  & -2&8915745669$\times 10^{-6}$ & 1&4951973738$\times 10^{-6}$ & 1&215325266$\times 10^{-5}$ \\
I+loop  & -3&7$\times 10^{-17}$ & 3&2$\times 10^{-18}$ & 1&284797177$\times 10^{-5}$ \\
\hline
& \multicolumn{2}{c}{ }
& \multicolumn{2}{c}{ $uc \to uc$}
& \multicolumn{2}{c}{ }
\\
\hline
I operator & 5&3681641565$\times 10^{-7}$ & 1&6010109373$\times 10^{-7}$ & -1&216280481$\times 10^{-7}$ \\
loop  & -5&3681641568$\times 10^{-7}$  & -1&6010109374$\times 10^{-7}$ & 3&231794702$\times 10^{-6}$ \\
I+loop  & -3&2$\times 10^{-17}$ & -1&7$\times 10^{-17}$ & 3&110166654$\times 10^{-6}$ \\
\hline
& \multicolumn{2}{c}{ }
& \multicolumn{2}{c}{ $ud \to ud$}
& \multicolumn{2}{c}{ }
\\
\hline
I operator & 1&4782975725$\times 10^{-6}$ & 4&4089012802$\times 10^{-7}$ & -3&349421572$\times 10^{-7}$ \\
loop  & -1&4782975726$\times 10^{-6}$  & -4&4089012788$\times 10^{-7}$ & 6&305527891$\times 10^{-6}$ \\
I+loop  & -3&7$\times 10^{-17}$ & 1&4$\times 10^{-16}$ & 5&970585734$\times 10^{-6}$ \\
\hline
& \multicolumn{2}{c}{ }
& \multicolumn{2}{c}{ $dd \to dd$}
& \multicolumn{2}{c}{ }
\\
\hline
I operator & 4&4102124035$\times 10^{-6}$ & -2&278825350$\times 10^{-6}$ & 1&0675989830$\times 10^{-6}$ \\
loop  & -4&4102124036$\times 10^{-6}$  & 2&278825350$\times 10^{-6}$ & 1&8933497562$\times 10^{-5}$ \\
I+loop  & -8&2$\times 10^{-17}$ & -1&1$\times 10^{-16}$ & 2&0001096545$\times 10^{-5}$ \\
\hline
& \multicolumn{2}{c}{ }
& \multicolumn{2}{c}{ $ds \to ds$}
& \multicolumn{2}{c}{ }
\\
\hline
I operator & 8&186414665$\times 10^{-7}$ & 2&4415310398$\times 10^{-7}$ & -1&854819651$\times 10^{-7}$ \\
loop  & -8&186414665$\times 10^{-7}$  & -2&4415310395$\times 10^{-7}$ & 5&284395333$\times 10^{-6}$ \\
I+loop  & -7&3$\times 10^{-17}$ & 2&7$\times 10^{-17}$ & 5&098913368$\times 10^{-6}$ \\
\hline
& \multicolumn{2}{c}{ }
& \multicolumn{2}{c}{ $gg \to \bar{u}u$}
& \multicolumn{2}{c}{ }
\\
\hline
I operator & 1&3039060448$\times 10^{-8}$ & -9&9737377238$\times 10^{-9}$ & 8&78497860$\times 10^{-11}$ \\
loop  & -1&3039060448$\times 10^{-8}$  & 9&9737377219$\times 10^{-9}$ & 3&21106128$\times 10^{-9}$ \\
I+loop  & -3&0$\times 10^{-20}$ & -2&0$\times 10^{-18}$ & 3&29891107$\times 10^{-9}$ \\
\hline
& \multicolumn{2}{c}{ }
& \multicolumn{2}{c}{ $gg \to \bar{d}d$}
& \multicolumn{2}{c}{ }
\\
\hline
I operator & 1&53496729122$\times 10^{-8}$ & -1&22781617391$\times 10^{-8}$ & 1&05371064$\times 10^{-9}$ \\
loop  & -1&53496729122$\times 10^{-8}$  & 1&22781617377$\times 10^{-8}$ & 2&17252400$\times 10^{-9}$ \\
I+loop  & -3&4$\times 10^{-20}$ & -1&4$\times 10^{-18}$ & 3&22623464$\times 10^{-9}$ \\
\hline
\end{tabular}\end{center}
 \end{footnotesize}
\end{table}

In the following, we provide the squared amplitude averaged over the initial-state 
helicities and colors. We also set $\alpha = \alpha_s = 1$ for simplicity. 
The top quark is decoupled from the running of $\alpha_s$, but its contribution 
is explicitly included in the one-loop amplitudes. All contributions including UV 
counterterms or a closed-quark loop with gluons attached to it are taken into account. 
As specified in \sect{sec:calculation}, diagrams including a closed-quark loop with the 
$Z/\gamma^*$ or/and the Higgs boson directly attached to it are excluded~\footnote{If 
the reader is interested in the discarded closed-quark loop contributions, we can provide these 
results upon request.}.
At tree level, we get
\begin{align}
  \overline{|\mathcal{A}_\text{LO}^{uu\rightarrow uu}|}^2 &= 3.40655603126\times 10^{-6},\nonumber \\
  \overline{|\mathcal{A}_\text{LO}^{uc\rightarrow uc}|}^2 &= 6.3242194040\times 10^{-7},\nonumber \\
  \overline{|\mathcal{A}_\text{LO}^{ud\rightarrow ud}|}^2 &= 1.741578298\times 10^{-6},\nonumber \\
  \overline{|\mathcal{A}_\text{LO}^{dd\rightarrow dd}|}^2 &= 5.195659083\times 10^{-6},\nonumber \\
  \overline{|\mathcal{A}_\text{LO}^{ds\rightarrow ds}|}^2 &= 9.644392564\times 10^{-7},\nonumber \\  
  \overline{|\mathcal{A}_\text{LO}^{gg\rightarrow \bar{u}u}|}^2 &= 9.4530961185\times 10^{-9},\nonumber \\
  \overline{|\mathcal{A}_\text{LO}^{gg\rightarrow \bar{d}d}|}^2 &= 1.11282506898\times 10^{-8}.
\end{align}
The interference amplitudes
$2\text{Re}(\mathcal{A}_\text{NLO}\mathcal{A}^{*}_\text{LO})$,
for the one-loop corrections  
and the I-operator contribution as defined in \bib{Catani:1996vz},
are given in \tab{table_PSP_QCD_jjjj}. For the one-loop integrals, we use the 
convention 
\bea
T_0 = \frac{\mu_R^{2\epsilon}\Gamma(1-\epsilon)}{i\pi^{2-\epsilon}}\int d^D q \frac{1}{(q^2 - m_1^2 + i0)\cdots},
\eea
with $D=4-2\epsilon$.
This amounts to dropping a factor ${(4\pi)^\epsilon}/{\Gamma(1-\epsilon)}$
both in the virtual corrections and the I-operator.
In addition, the conventional dimensional regularization method~\cite{'tHooft:1972fi}
with $\mu_{R} = M_Z$ is used. 
Switching from the conventional dimensional regularization to dimensional reduction method 
induces a finite shift. 
This shift can be calculated observing that 
the sum $|\mathcal{A}_\text{LO}|^2+2\text{Re}(\mathcal{A}_\text{NLO}\mathcal{A}^{*}_\text{LO})$ 
must remain unchanged~\cite{Catani:1996pk}. 
The shift 
on the Born amplitude squared is given by the following change in the strong coupling constant, 
see e.g. \bib{Kunszt:1993sd},
\bea
\alpha_s^{\overline{DR}} = \alpha_s^{\overline{MS}}\left(1+\frac{\alpha_s}{4\pi}\right).
\eea
The shift on the I-operator contribution can easily be obtained using the rule given in \bib{Catani:1996vz}. 

\acknowledgments
We would like to thank Michael Rauch for help and useful discussions.
LDN and DZ are supported in part by the Deutsche Forschungsgemeinschaft
via the Sonderforschungsbereich/Transregio SFB/TR-9 ``Computational
Particle Physics''. FC acknowledges financial support from the Marie Curie
Actions (PIEF-GA-2011-298960), by the LHCPhenonet (PITN-GA-2010-264564) and by
the MINECO (FPA2011-23596). MK is funded by the
Graduiertenkolleg 1694 ``Elementarteilchenphysik bei h\"ochster
Energie und h\"ochster Pr\"azision''.


\begin{thebibliography}{10}

\bibitem{ATLAS001}
T.~A. collaboration, {\it {Evidence for electroweak production of
  $W^{\pm}W^{\pm}jj$ in $pp$ collisions at $\sqrt{s}=8$ TeV with the ATLAS
  detector}}, .

\bibitem{Jager:2009xx}
B.~Jager, C.~Oleari, and D.~Zeppenfeld, {\it {Next-to-leading order QCD
  corrections to $W^+ W^+ jj$ and $W^- W^- jj$ production via weak-boson
  fusion}},  {\em Phys.Rev.} {\bf D80} (2009) 034022,
  [\href{http://xxx.lanl.gov/abs/0907.0580}{{\tt 0907.0580}}].

\bibitem{Denner:2012dz}
A.~Denner, L.~Hosekova, and S.~Kallweit, {\it {NLO QCD corrections to $W^+ W^+
  jj$ production in vector-boson fusion at the LHC}},  {\em Phys.Rev.} {\bf
  D86} (2012) 114014, [\href{http://xxx.lanl.gov/abs/1209.2389}{{\tt
  1209.2389}}].

\bibitem{Melia:2010bm}
T.~Melia, K.~Melnikov, R.~Rontsch, and G.~Zanderighi, {\it {Next-to-leading
  order QCD predictions for $W^+W^+jj$ production at the LHC}},  {\em JHEP}
  {\bf 1012} (2010) 053, [\href{http://xxx.lanl.gov/abs/1007.5313}{{\tt
  1007.5313}}].

\bibitem{Campanario:2013gea}
F.~Campanario, M.~Kerner, L.~D. Ninh, and D.~Zeppenfeld, {\it {Next-to-leading
  order QCD corrections to $W^+W^+$ and $W^-W^-$ production in association with
  two jets}},  {\em Phys.Rev.} {\bf D89} (2014) 054009,
  [\href{http://xxx.lanl.gov/abs/1311.6738}{{\tt arXiv:1311.6738}}].

\bibitem{Gleisberg:2008ta}
T.~Gleisberg, S.~Hoeche, F.~Krauss, M.~Schonherr, S.~Schumann, et~al., {\it
  {Event generation with SHERPA 1.1}},  {\em JHEP} {\bf 0902} (2009) 007,
  [\href{http://xxx.lanl.gov/abs/0811.4622}{{\tt 0811.4622}}].

\bibitem{Jager:2006cp}
B.~Jager, C.~Oleari, and D.~Zeppenfeld, {\it {Next-to-leading order QCD
  corrections to Z boson pair production via vector-boson fusion}},  {\em
  Phys.Rev.} {\bf D73} (2006) 113006,
  [\href{http://xxx.lanl.gov/abs/hep-ph/0604200}{{\tt hep-ph/0604200}}].

\bibitem{Melia:2011dw}
T.~Melia, K.~Melnikov, R.~Rontsch, and G.~Zanderighi, {\it {NLO QCD corrections
  for $W^+W^-$ pair production in association with two jets at hadron
  colliders}},  {\em Phys.Rev.} {\bf D83} (2011) 114043,
  [\href{http://xxx.lanl.gov/abs/1104.2327}{{\tt 1104.2327}}].

\bibitem{Greiner:2012im}
N.~Greiner, G.~Heinrich, P.~Mastrolia, G.~Ossola, T.~Reiter, et~al., {\it {NLO
  QCD corrections to the production of $W^+ W^-$ plus two jets at the LHC}},
  {\em Phys.Lett.} {\bf B713} (2012) 277--283,
  [\href{http://xxx.lanl.gov/abs/1202.6004}{{\tt 1202.6004}}].

\bibitem{Campanario:2013qba}
F.~Campanario, M.~Kerner, L.~D. Ninh, and D.~Zeppenfeld, {\it {WZ production in
  association with two jets at NLO in QCD}},  {\em Phys. Rev. Lett.} {\bf 111}
  (2013) 052003, [\href{http://xxx.lanl.gov/abs/1305.1623}{{\tt 1305.1623}}].

\bibitem{Gehrmann:2013bga}
T.~Gehrmann, N.~Greiner, and G.~Heinrich, {\it {Precise QCD predictions for the
  production of a photon pair in association with two jets}},  {\em
  Phys.Rev.Lett.} {\bf 111} (2013) 222002,
  [\href{http://xxx.lanl.gov/abs/1308.3660}{{\tt arXiv:1308.3660}}].

\bibitem{Badger:2013ava}
S.~Badger, A.~Guffanti, and V.~Yundin, {\it {Next-to-leading order QCD
  corrections to di-photon production in association with up to three jets at
  the Large Hadron Collider}},  {\em JHEP} {\bf 1403} (2014) 122,
  [\href{http://xxx.lanl.gov/abs/1312.5927}{{\tt arXiv:1312.5927}}].

\bibitem{Bern:2014vza}
Z.~Bern, L.~Dixon, F.~Febres~Cordero, S.~Hoeche, H.~Ita, et~al., {\it
  {Next-to-Leading Order Gamma Gamma + 2-Jet Production at the LHC}},
  \href{http://xxx.lanl.gov/abs/1402.4127}{{\tt arXiv:1402.4127}}.

\bibitem{Campanario:2014dpa}
F.~Campanario, M.~Kerner, L.~D. Ninh, and D.~Zeppenfeld, {\it {Next-to-leading
  order QCD corrections to $W \gamma$ production in association with two
  jets}},  \href{http://xxx.lanl.gov/abs/1402.0505}{{\tt arXiv:1402.0505}}.

\bibitem{Alwall:2014hca}
J.~Alwall, R.~Frederix, S.~Frixione, V.~Hirschi, F.~Maltoni, et~al., {\it {The
  automated computation of tree-level and next-to-leading order differential
  cross sections, and their matching to parton shower simulations}},
  \href{http://xxx.lanl.gov/abs/1405.0301}{{\tt arXiv:1405.0301}}.

\bibitem{Arnold:2008rz}
K.~Arnold, M.~Bahr, G.~Bozzi, F.~Campanario, C.~Englert, et~al., {\it {VBFNLO:
  A Parton level Monte Carlo for processes with electroweak bosons}},  {\em
  Comput.Phys.Commun.} {\bf 180} (2009) 1661--1670,
  [\href{http://xxx.lanl.gov/abs/0811.4559}{{\tt 0811.4559}}].

\bibitem{Baglio:2014uba}
J.~Baglio, J.~Bellm, F.~Campanario, B.~Feigl, J.~Frank, et~al., {\it {Release
  Note - VBFNLO 2.7.0}},  \href{http://xxx.lanl.gov/abs/1404.3940}{{\tt
  arXiv:1404.3940}}.

\bibitem{'tHooft:1972fi}
G.~'t~Hooft and M.~Veltman, {\it {Regularization and Renormalization of Gauge
  Fields}},  {\em Nucl.Phys.} {\bf B44} (1972) 189--213.

\bibitem{Chanowitz:1979zu}
M.~S. Chanowitz, M.~Furman, and I.~Hinchliffe, {\it {The Axial Current in
  Dimensional Regularization}},  {\em Nucl.Phys.} {\bf B159} (1979) 225.

\bibitem{Catani:1996vz}
S.~Catani and M.~Seymour, {\it {A General algorithm for calculating jet
  cross-sections in NLO QCD}},  {\em Nucl.Phys.} {\bf B485} (1997) 291--419,
  [\href{http://xxx.lanl.gov/abs/hep-ph/9605323}{{\tt hep-ph/9605323}}].

\bibitem{Campanario:2012bh}
F.~Campanario, Q.~Li, M.~Rauch, and M.~Spira, {\it {ZZ+jet production via gluon
  fusion at the LHC}},  {\em JHEP} {\bf 1306} (2013) 069,
  [\href{http://xxx.lanl.gov/abs/1211.5429}{{\tt arXiv:1211.5429}}].

\bibitem{'tHooft:1978xw}
G.~'t~Hooft and M.~Veltman, {\it {Scalar One Loop Integrals}},  {\em
  Nucl.Phys.} {\bf B153} (1979) 365--401.

\bibitem{Bern:1993kr}
Z.~Bern, L.~J. Dixon, and D.~A. Kosower, {\it {Dimensionally regulated pentagon
  integrals}},  {\em Nucl.Phys.} {\bf B412} (1994) 751--816,
  [\href{http://xxx.lanl.gov/abs/hep-ph/9306240}{{\tt hep-ph/9306240}}].

\bibitem{Dittmaier:2003bc}
S.~Dittmaier, {\it {Separation of soft and collinear singularities from one
  loop N point integrals}},  {\em Nucl.Phys.} {\bf B675} (2003) 447--466,
  [\href{http://xxx.lanl.gov/abs/hep-ph/0308246}{{\tt hep-ph/0308246}}].

\bibitem{Nhung:2009pm}
D.~T. Nhung and L.~D. Ninh, {\it {D0C : A code to calculate scalar one-loop
  four-point integrals with complex masses}},  {\em Comput. Phys. Commun.} {\bf
  180} (2009) 2258--2267, [\href{http://xxx.lanl.gov/abs/0902.0325}{{\tt
  0902.0325}}].

\bibitem{Denner:2010tr}
A.~Denner and S.~Dittmaier, {\it {Scalar one-loop 4-point integrals}},  {\em
  Nucl.Phys.} {\bf B844} (2011) 199--242,
  [\href{http://xxx.lanl.gov/abs/1005.2076}{{\tt 1005.2076}}].

\bibitem{Passarino:1978jh}
G.~Passarino and M.~Veltman, {\it {One Loop Corrections for $e^+ e^-$
  Annihilation Into $\mu^+ \mu^-$ in the Weinberg Model}},  {\em Nucl.Phys.}
  {\bf B160} (1979) 151.

\bibitem{Denner:2005nn}
A.~Denner and S.~Dittmaier, {\it {Reduction schemes for one-loop tensor
  integrals}},  {\em Nucl.Phys.} {\bf B734} (2006) 62--115,
  [\href{http://xxx.lanl.gov/abs/hep-ph/0509141}{{\tt hep-ph/0509141}}].

\bibitem{Campanario:2011cs}
F.~Campanario, {\it {Towards $pp \to VVjj$ at NLO QCD: Bosonic contributions to
  triple vector boson production plus jet}},  {\em JHEP} {\bf 1110} (2011) 070,
  [\href{http://xxx.lanl.gov/abs/1105.0920}{{\tt 1105.0920}}].

\bibitem{Binoth:2005ff}
T.~Binoth, J.~P. Guillet, G.~Heinrich, E.~Pilon, and C.~Schubert, {\it {An
  Algebraic/numerical formalism for one-loop multi-leg amplitudes}},  {\em
  JHEP} {\bf 0510} (2005) 015,
  [\href{http://xxx.lanl.gov/abs/hep-ph/0504267}{{\tt hep-ph/0504267}}].

\bibitem{Gleisberg:2008fv}
T.~Gleisberg and S.~Hoeche, {\it {Comix, a new matrix element generator}},
  {\em JHEP} {\bf 0812} (2008) 039,
  [\href{http://xxx.lanl.gov/abs/0808.3674}{{\tt 0808.3674}}].

\bibitem{Hahn:2000kx}
T.~Hahn, {\it {Generating Feynman diagrams and amplitudes with FeynArts 3}},
  {\em Comput.Phys.Commun.} {\bf 140} (2001) 418--431,
  [\href{http://xxx.lanl.gov/abs/hep-ph/0012260}{{\tt hep-ph/0012260}}].

\bibitem{Hahn:1998yk}
T.~Hahn and M.~Perez-Victoria, {\it {Automatized one-loop calculations in four
  and D dimensions}},  {\em Comput. Phys. Commun.} {\bf 118} (1999) 153--165,
  [\href{http://xxx.lanl.gov/abs/hep-ph/9807565}{{\tt hep-ph/9807565}}].

\bibitem{Martin:2009iq}
A.~Martin, W.~Stirling, R.~Thorne, and G.~Watt, {\it {Parton distributions for
  the LHC}},  {\em Eur.Phys.J.} {\bf C63} (2009) 189--285,
  [\href{http://xxx.lanl.gov/abs/0901.0002}{{\tt 0901.0002}}].

\bibitem{Cacciari:2008gp}
M.~Cacciari, G.~P. Salam, and G.~Soyez, {\it {The Anti-k(t) jet clustering
  algorithm}},  {\em JHEP} {\bf 0804} (2008) 063,
  [\href{http://xxx.lanl.gov/abs/0802.1189}{{\tt 0802.1189}}].

\bibitem{Catani:1996pk}
S.~Catani, M.~Seymour, and Z.~Trocsanyi, {\it {Regularization scheme
  independence and unitarity in QCD cross-sections}},  {\em Phys.Rev.} {\bf
  D55} (1997) 6819--6829, [\href{http://xxx.lanl.gov/abs/hep-ph/9610553}{{\tt
  hep-ph/9610553}}].

\bibitem{Kunszt:1993sd}
Z.~Kunszt, A.~Signer, and Z.~Trocsanyi, {\it {One loop helicity amplitudes for
  all $2 \to 2$ processes in QCD and $N=1$ supersymmetric Yang-Mills theory}},
  {\em Nucl.Phys.} {\bf B411} (1994) 397--442,
  [\href{http://xxx.lanl.gov/abs/hep-ph/9305239}{{\tt hep-ph/9305239}}].

\end{thebibliography}
\providecommand{\href}[2]{#2}\begingroup\raggedright\endgroup

\end{document}